# Opto-Mechanical Interactions in Multi-Core Optical Fibers and Their Applications

Hilel Hagai Diamandi, Yosef London, Arik Bergman, *Member, IEEE*, Gil Bashan, Javier Madrigal, David Barrera, Salvador Sales, *Senior Member, IEEE*, and Avi Zadok

*(Invited Paper)*

*Abstract*—Optical fibers containing multiple cores are being developed towards capacity enhancement in space-division multiplexed optical communication networks. In many cases, the fibers are designed for negligible direct coupling of optical power among the cores. The cores remain, however, embedded in a single, mechanically-unified cladding. Elastic (or acoustic) modes supported by the fiber cladding geometry are in overlap with multiple cores. Acoustic waves may be stimulated by light in any core through electrostriction. Once excited, the acoustic waves may induce photo-elastic perturbations to optical waves in other cores as well. Such opto-mechanical coupling gives rise to inter-core cross-phase modulation effects, even when direct optical crosstalk is very weak. The cross-phase modulation spectrum reaches hundreds of megahertz frequencies. It may consist of discrete and narrow peaks, or may become quasi-continuous, depending on the geometric layout. The magnitude of the effect at the resonance frequencies is comparable with that of intra-core cross-phase modulation due to Kerr nonlinearity. Two potential applications are demonstrated: single-frequency opto-electronic oscillators that do not require radio-frequency electrical filters, and point-sensing of liquids outside the cladding of multi-core fibers, where light cannot reach.

Manuscript received October 2, 2019; revised December 8, 2019; accepted December 8, 2019. Date of publication December 11, 2019; date of current version January 14, 2020. This work was supported in part by a Starter Grant from the European Research Council (ERC) under Grant H2020-ERC-2015-STG 679228 (L-SID), in part by the Israeli Ministry of Science and Technology under Grant 61047, and in part by the Spanish Ministry of Economy and Competitiveness under the DIMENSION TEC2017 88029-R Project. H. H. Diamandi was supported by the Azrieli Foundation for the award of an Azrieli Fellowship. The work of J. Madrigal was supported by Universitat Politècnica de València scholarship PAID-01-18. The work of D. Barrera was supported by Spanish MICINN fellowship IJCI-2017-32476. *(Corresponding author: Avi Zadok.)*

H. H. Diamandi, Y. London, G. Bashan, and A. Zadok are with the Faculty of Engineering and Institute for Nano-Technology and Advanced Materials, Bar-Ilan University, Ramat-Gan 5290002, Israel (e-mail: hagaid@gmail.com; yoseflondon@gmail.com; gillbashan@gmail.com; Avinoam.Zadok@biu.ac.il).

A. Bergman is with the Faculty of Engineering and Institute for Nano-Technology and Advanced Materials, Bar-Ilan University, Ramat-Gan 5290002, Israel, and also with Advanced Science Research Center, City University of New York, New York, NY 10031 USA (e-mail: bergman.arik@gmail.com).

J. Madrigal and S. Sales are with Photonics Research Labs, iTEAM Research Institute, Universitat Politècnica de València, 46022 Valencia, Spain (e-mail: jamadmad@iteam.upv.es; ssales@dcom.upv.es).

D. Barrera is with the Department of Electronics, University of Alcalá, 28805 Alcalá de Henares, Spain (e-mail: david.barrera@uah.es).

Color versions of one or more of the figures in this article are available online at http://ieeexplore.ieee.org.

Digital Object Identifier 10.1109/JSTQE.2019.2958933

*Index Terms*—Opto-mechanics, multi-core fibers, stimulated Brillouin scattering, opto-electronic oscillators, optical fiber sensors, nonlinear fiber-optics.

## I. INTRODUCTION

THE capacity of optical communication over single-mode fibers is reaching its fundamental limits [1], [2]. The leading solution for capacity enhancement is the multiplexing of independent data channels on different spatial guided modes of a single fiber, and their separation at the receiver end [3], [4]. A common realization of such space-division multiplexing networks makes use of fibers with multiple cores [5]. In addition to data communication, multi-core fibers (MCFs) also find applications in distributed shape sensing [6], [7], microwave-photonics [8], [9], opto-electronic oscillators [10], and fiber lasers [11].

In many cases, the cross-sections of MCFs are designed with large separation among the constituent cores [5]. The direct coupling of optical power among the cores in such designs is negligible. The absence of coupling simplifies many communication, sensing and signal processing applications. However, the cores remain assembled as part of a single, mechanically-unified cladding. The cladding of an optical fiber supports several classes of guided acoustic modes [12]–[17]. Unlike the optical modes, which are largely confined to core regions, the transverse profiles of guided acoustic modes typically span the entire cross-section of the cladding [12]–[17] and may overlap with several cores.

The oscillation of acoustic modes may by stimulated by guided optical waves through the mechanism of electrostriction [13]–[17]. The acoustic waves, in turn, may scatter and modulate light via photo-elasticity [13]–[17]. Interactions between guided light and sound waves were first studied in standard single-mode fibers in 1985 [13], [17], and have been thoroughly investigated in micro-structured and photonic-crystal fibers as well [18]–[20]. The effect has also been demonstrated in integrated-photonic waveguides [21]–[23]. However, it was seldom considered in MCFs until recently. In one example, switching among cores was proposed using piezo-electric stimulation of flexural acoustic waves [24].

We have been investigating opto-mechanical interactions in commercially-available MCFs over the last three years [25]. The coupling of optical and acoustic modes in MCFs presents a unique property: Due to the broad transverse profiles of the guided acoustic modes, sound waves that are stimulated by light in any core may induce photo-elastic index perturbations in other cores as well [25]. These perturbations give rise to inter-core, opto-mechanical cross-phase modulation (XPM), which takes place even in the absence of direct optical coupling [25]. The frequencies of inter-core XPM range from tens to hundreds of megahertz. The modulation spectra may consist of discrete and narrow peaks, or may become quasi-continuous, depending on the geometry of the fiber cross-section and choice of cores [25]. The magnitude of the effect at resonance frequencies is comparable with that of intra-core XPM due to Kerr nonlinearity. Similar coupling between integrated-photonic waveguides that are parts of a common silicon membrane has been studied independently [26], [27].

The analysis of opto-mechanical coupling through radial guided acoustic modes in MCFs is presented in Section II. Analytic expressions for the magnitude and spectrum of inter-core XPM are provided, and corresponding measurements are reported as well. Two applications of the effects are demonstrated in Sections III and IV, respectively: single-frequency opto-electronic oscillators that do not require inline radio-frequency electrical filters, and point-sensing of liquids outside the MCF cladding where light cannot reach. These applications illustrate the potential of interactions among the plurality of optical and acoustic modes that are supported by MCFs. Conclusions are drawn in Section V. The description below is intended primarily for readers who are familiar with fiber-optics, but may not be specialists in opto-mechanics. Parts of the results appear in separate, earlier publications [25], [28], however the reader may benefit from the more unified perspective provided below. Some results were presented only briefly in a recent conference [29].

## II. OPTO-MECHANICAL INTERACTIONS IN MULTI-CORE FIBERS

The cylindrical cladding of optical fibers supports guided acoustic modes that propagate in the axial direction. The most general modes are torsional-radial: The transverse displacement vectors of the fiber medium in the oscillation of these modes include both radial and azimuthal components, and the local magnitude of each depends on both the radial and azimuthal coordinates [12]–[16]. Light propagating in an off-axis core of an MCF may stimulate many torsional-radial modes [25]. For simplicity, however, the following discussion is restricted for the most part to radial acoustic modes, in which displacement is in the radial direction only and its magnitude depends only on the radial coordinate $r$ [12]. We refer to the discrete set of radial guided acoustic modes as $R_{0,m}$, where $m$ is a positive integer. Inter-core XPM is restricted to the contributions of radial acoustic modes only when pump light is propagating at the inner, on-axis core and the polarization of the pump wave is scrambled (see also later in this section). A discussion of the more general case of opto-mechanical interactions involving torsional-radial modes may be found in [25].

Consider an MCF of silica cladding with a radius $a$, which is surrounded by air. Throughout the following analysis,

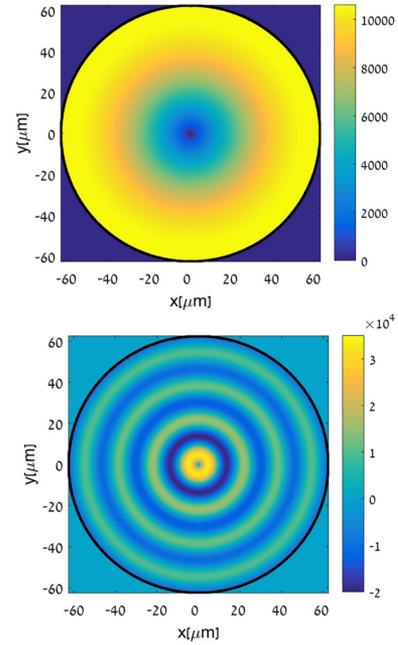

Fig. 1. Normalized transverse displacement profiles of radial acoustic modes that are guided by the cladding of an optical fiber. The cladding diameter is 125 $\mu$m. Top: radial mode $R_{0,1}$. Bottom: radial mode $R_{0,8}$.

we disregard the small-scale differences between the elastic properties of cores and cladding regions, and assume that the fiber cross-section is mechanically uniform. The normalized radial displacement profile of mode $R_{0,m}$ is given by [12]–[16]:

$$u_r^{(m)}(r) = J_1\left[(\Omega_m r)/v_d\right] \bigg/ \sqrt{2\pi \int_0^a \{J_1\left[(\Omega_m r)/v_d\right]\}^2 r\,\mathrm{d}r}. \tag{1}$$

In Eq. (1), $J_1$ denotes the first-order Bessel function of the first kind, $v_d$ is the velocity of dilatational acoustic waves in silica, and $\Omega_m$ is the cut-off angular frequency of mode $R_{0,m}$ (to be discussed further shortly). The units of the displacement profile are m$^{-1}$, and it is normalized so that $2\pi \int_0^a |u_r^{(m)}|^2 r\,\mathrm{d}r = 1$.

The boundary condition at $r = a$ for a fiber in air requires that the radial component of the stress tensor should equal zero [12]–[16]. This requirement can be brought to the following form [12]–[16]:

$$\left(1 - \frac{v_s^2}{v_d^2}\right) J_0(\xi) = \frac{v_s^2}{v_d^2} J_2(\xi). \tag{2}$$

Here $J_0$ and $J_2$ are the zero-order and second-order Bessel functions of the first kind, respectively, and $v_s$ represents the velocity of acoustic shear waves in silica. Equation (2) has a discrete set of solutions $\{\xi_m\}$. The cut-off angular frequency of mode $R_{0,m}$ is given by $\Omega_m = (v_d/a)\xi_m$. With knowledge of $\Omega_m$, the transverse profile of the radial mode is fully determined. As examples, Fig. 1 shows $u_r^{(1)}(r)$ and $u_r^{(8)}(r)$. The cut-off angular frequencies of the two modes for $a = 62.5$ $\mu$m are $2\pi \times 30$ MHz and $2\pi \times 370$ MHz, respectively.

At angular frequencies of oscillations $\Omega$ that are close to cut-off, the dispersion relation between angular frequency and axial

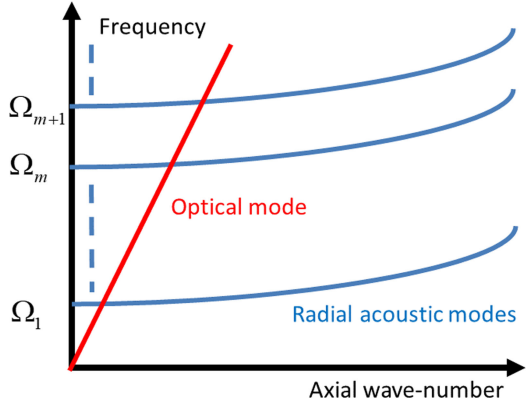

Fig. 2. Schematic illustration of the dispersion relations between angular frequency and axial wave-number of radial acoustic modes that are guided by an optical fiber (blue). The dispersion relation of the single optical mode in any of the fiber cores is shown in red.

wave-number $k_z^{(m)}$ in mode $R_{0,m}$ is approximately given by [12]:

$$\Omega = \sqrt{\left(k_z^{(m)} v_d\right)^2 + \Omega_m^2}. \quad (3)$$

The dispersion relation is illustrated in Fig. 2. The axial wave-number vanishes for $\Omega \to \Omega_m$. The group velocity in the axial direction $\partial \Omega / \partial k_z^{(m)}$ approaches zero at cut-off. In contrast, the axial phase velocity $\Omega / k_z^{(m)}$ approaches infinity. In particular, very close to cut-off, the axial phase velocity may equal that of guided light (see Fig. 2). The agreement in velocities provides phase-matching in opto-mechanical interactions, as discussed later. Optically stimulated guided acoustic waves are almost entirely transverse: The axial component of their wave vector is orders of magnitude smaller than the transverse component. The axial acoustic wavelength is typically on the order of tens of centimeters.

Consider next an optical pump wave that is propagating at the central, on-axis core of an MCF. We denote the transverse profile of the pump optical field as $E_T^{(pump)}(r)$, in units of m$^{-1}$. It is normalized to $2\pi \int_0^a |E_T^{(pump)}(r)|^2 r dr = 1$. The instantaneous pump power (in W) at a given fiber position is $P(t)$, where $t$ stands for time. The radio-frequency Fourier transform of $P(t)$ is denoted by $\tilde{P}(\Omega)$ [W $\times$ rad$^{-1}$Hz$^{-1}$]. The propagation of the optical pump wave is associated with a transverse electrostrictive force per unit volume. Let us assume first, without loss of generality, that the pump wave is linearly polarized along the $\hat{x}$ axis. The direction of polarization breaks the radial symmetry of the electrostrictive driving force. It consists of a radial component $\tilde{F}_r$ and an azimuthal component $\tilde{F}_\varphi$, that depend on the azimuthal coordinate $\varphi$ as well as on $r$ [12]–[16]:

$$\tilde{F}_r(r, \Omega, \varphi) = -\frac{a_1 + 4a_2 + a_1 \cos(2\varphi)}{4nc}$$
$$\times E_T^{(pump)}(r) \frac{\partial E_T^{(pump)}}{\partial r}(r) \tilde{P}(\Omega), \quad (4)$$

and

$$\tilde{F}_\varphi(r, \Omega, \varphi) = \frac{a_1 \sin(2\varphi)}{4nc} E_T^{(pump)}(r) \frac{\partial E_T^{(pump)}}{\partial r}(r) \tilde{P}(\Omega). \quad (5)$$

In Eq. (4) and Eq. (5) $n$ is the refractive index of silica, $c$ denotes the speed of light in vacuum, and $a_1$ and $a_2$ are given by elements of the photo-elastic tensor $\mathbf{P}$ of silica: $a_1 = -n^4(P_{11} - P_{12}) = 0.66$, $a_2 = -n^4 P_{12} = -1.19$. The units of the terms $\tilde{F}_r$ and $\tilde{F}_\varphi$ are N $\times$ m$^{-3}$ $\times$ rad$^{-1}$Hz$^{-1}$. Due to its azimuthal dependence, the electro-strictive driving force may simulate the oscillations of non-radial guided acoustic waves as well. The azimuthal orientation of non-radial modes would follow that of the pump polarization. However, throughout this work, the state of polarization of the optical pump wave is deliberately scrambled, so that the effect of non-radial modes cancels out on average. The remaining electrostrictive force term of consequence is entirely radial:

$$\tilde{F}_r(r, \Omega) = -\frac{a_1 + 4a_2}{4nc} E_T^{(pump)}(r) \frac{\partial E_T^{(pump)}}{\partial r}(r) \tilde{P}(\Omega). \quad (6)$$

The electrostrictive force serves as a driving term in the elastic wave equation [12]–[16], and may stimulate the oscillations of guided radial acoustic modes $R_{0,m}$. The local magnitude of material displacement in each mode is given by $\tilde{U}^{(m)}(\Omega, r) = \tilde{A}_r^{(m)}(\Omega) u_r^{(m)}(r)$, where:

$$\tilde{A}_r^{(m)}(\Omega) \approx \frac{1}{4nc\rho_0 \Gamma_m \Omega_m} Q_{ES}^{(m)} \left[\frac{\tilde{P}(\Omega)}{j - 2(\Omega - \Omega_m)/\Gamma_m}\right]. \quad (7)$$

The units of $\tilde{A}_r^{(m)}(\Omega)$ are [m$^2$ $\times$ rad$^{-1}$Hz$^{-1}$]. Here $\rho_0$ is the density of silica, and $\Gamma_m$ denote the linewidths of modal oscillations which also signify their decay rates. Also in Eq. (7), we have defined the transverse overlap integral of electrostrictive stimulation:

$$Q_{ES}^{(m)} \equiv -(a_1 + 4a_2)$$
$$\cdot 2\pi \int_0^a E_T^{(pump)}(r) \frac{dE_T^{(pump)}(r)}{dr} u_r^{(m)}(r) r dr \quad (8)$$

The linewidths $\Gamma_m$ consist of two contributions: an intrinsic component that is associated with acoustic losses in silica, and a boundary-related term that quantifies the dissipation of acoustic energy to the surrounding medium [30]. The intrinsic linewidth is frequency dependent, and it is on the order of 100 kHz for acoustic waves at hundreds of megahertz frequencies [30]. The external dissipation contribution is determined by the mechanical impedance of the medium outside the cladding [30]. Measurements of that linewidth component are therefore central to new optical fiber sensing concepts that are based on opto-mechanics [30], and will be discussed in greater detail in Section IV.

Equation (7) highlights the two conditions that are necessary for electrostrictive stimulation of radial guided acoustic modes. First, phase matching between the axial and temporal properties of the optical and acoustic modes must be achieved. Such matching is optimal near the cut-off angular frequency $\Omega_m$ as discussed above. This condition manifests in the Lorentzian frequency dependence of the modal displacement magnitudes. A second requirement is an agreement between the transverse profiles of the electrostrictive driving force and the acoustic displacement, as expressed in the spatial overlap integral $Q_{ES}^{(m)}$. The driving force is largely confined to the on-axis core. Referring to Fig. 1, for example, one may expect that the electrostrictive stimulation of mode $R_{0,8}$ would be more efficient than that of mode $R_{0,1}$.

Equation (7) also states that in order to stimulate a guided acoustic wave, the instantaneous power of the optical pump field must oscillate at angular frequency $\Omega$. The pump wave should therefore consist of at least two spectral components that are detuned in angular frequency by $\Omega$. The two components are co-propagating in the same direction along the fiber. This property is in marked contrast with the more widely known phenomenon of backwards stimulated Brillouin scattering (backwards SBS) in optical fibers [31]. Backwards SBS is also driven by electrostriction [31]. In that case, however, longitudinal acoustic waves are stimulated by two optical field components that are counter-propagating. The frequencies of the acoustic waves in backwards SBS are typically on the order of 10 GHz, and their transverse profiles are largely confined to the core just like the optical mode. In order to note the similarities and distinctions between the two effects, the stimulation of transverse guided acoustic waves is often referred to as forward SBS [15].

The material displacement of the acoustic wave is associated with a tensor of strain perturbations that oscillate at angular frequency $\Omega$. The magnitude of the local normal strain in the radial direction is given by $\tilde{S}_{rr}^{(m)}(r,\Omega) = \tilde{A}_r^{(m)}(\Omega) du_r^{(m)}(r)/dr$, and that of normal strain in the azimuthal direction equals $\tilde{S}_{\varphi\varphi}^{(m)}(r,\Omega) = \tilde{A}_r^{(m)}(\Omega) u_r^{(m)}(r)/r$ [13]–[16]. The radial guided acoustic modes do not induce shear strain. Strain in the material, in turn, gives rise to perturbations in the local values of the dielectric tensor through the photo-elastic effect. The magnitudes of photo-elastic perturbations to the radial and azimuthal terms of the local dielectric tensor are given by [13]–[16]:

$$\delta\tilde{\varepsilon}_{rr}^{(m)}(r,\Omega) = (a_1 + a_2)\tilde{S}_{rr}^{(m)}(r,\Omega) + a_2 \tilde{S}_{\varphi\varphi}^{(m)}(r,\Omega), \quad (9)$$

and:

$$\delta\tilde{\varepsilon}_{\varphi\varphi}^{(m)}(r,\Omega) = (a_1 + a_2)\tilde{S}_{\varphi\varphi}^{(m)}(r,\Omega) + a_2 \tilde{S}_{rr}^{(m)}(r,\Omega). \quad (10)$$

Calculations of the two terms for mode $R_{0,8}$ are shown in Fig. 3.

In this work, we are primarily interested in the effect of the acoustically-induced perturbations to the dielectric tensor on the propagation of probe light in an outer, off-axis core of the MCF. We denote the transverse position of the center of the outer core as $\vec{r}_{probe}$. Let $\hat{e}_1$ represent a unit vector in the direction of $\vec{r}_{probe}$, and $\hat{e}_2$ represent the orthogonal unit vector. The normalized transverse profile of the probe field is noted by $E_T^{(probe)}(r,\varphi)$, with normalization $\int_0^a \int_0^{2\pi} |E_T^{(probe)}(r,\varphi)|^2 r dr d\varphi = 1$. If $|\vec{r}_{probe}|$ is much larger than the mode field diameters of both $E_T^{(pump)}(r)$ and

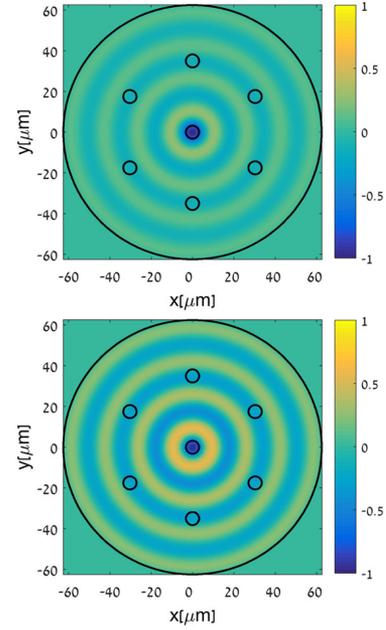

Fig. 3. Normalized transverse profiles of the photo-elastic perturbation magnitude to the dielectric tensor of silica, due to the oscillations of radial guided acoustic mode $R_{0,8}$. Top: radial component. Bottom: azimuthal component. The cladding diameter is 125 $\mu$m. The seven cores of the fiber used in this work are noted by black circles.

$E_T^{(probe)}(r,\varphi)$, the radial direction throughout the extent of the probe field is closely aligned with $\hat{e}_1$. Similarly, the azimuthal direction throughout $E_T^{(probe)}(r,\varphi)$ is approximately the same as $\hat{e}_2$.

Subject to the above assumption, we may distinguish between two states of polarization of the probe wave. If the probe field is polarized along $\hat{e}_1$, photo-elastic perturbations to the propagation of the probe wave are approximately determined by the radial element $\delta\tilde{\varepsilon}_{rr}^{(m)}(r,\Omega)$ of Eq. (9). If the probe field is polarized along $\hat{e}_2$ instead, we may estimate the photo-elastic perturbations based on $\delta\tilde{\varepsilon}_{\varphi\varphi}^{(m)}(r,\Omega)$ of Eq. (10). In the former case, the magnitude of perturbations to the effective index of the probe wave may be written as:

$$\overline{\delta\tilde{n}}^{(m,1)}(\Omega) = -\frac{1}{8n^2 c \rho_0 \Gamma_m \Omega_m} Q_{ES}^{(m)} Q_{PE}^{(m,1)}$$
$$\times \frac{\tilde{P}(\Omega)}{j + 2(\Omega - \Omega_m)/\Gamma_m}. \quad (11)$$

Here we have defined the photo-elastic spatial overlap integral $Q_{PE}^{(m,1)}$, which expresses the extent of agreement between the transverse profile of the probe optical mode and the radial element of the dielectric perturbations tensor:

$$Q_{PE}^{(m,1)} = \int_0^{2\pi} \int_0^a \left[(a_1 + a_2)\frac{du_r^{(m)}(r)}{dr} + a_2 \frac{u_r^{(m)}(r)}{r}\right]$$
$$\times \left|E_T^{(probe)}(r,\varphi)\right|^2 r dr d\varphi. \quad (12)$$

Similarly, we may express the magnitude of index perturbations for probe light polarized along $\hat{e}_2$:

$$\overline{\delta\tilde{n}}^{(m,2)}(\Omega) = -\frac{1}{8n^2 c\rho_0 \Gamma_m \Omega_m} Q_{ES}^{(m)} Q_{PE}^{(m,2)}$$
$$\times \frac{\tilde{P}(\Omega)}{j + 2(\Omega - \Omega_m)/\Gamma_m}, \quad (13)$$

with a second photo-elastic overlap integral $Q_{PE}^{(m,2)}$ that relates to the azimuthal element of the dielectric tensor:

$$Q_{PE}^{(m,2)} = \int_0^{2\pi}\int_0^a \left[a_2 \frac{du_r^{(m)}(r)}{dr} + (a_1+a_2)\frac{u_r^{(m)}(r)}{r}\right]$$
$$\times \left|E_T^{(probe)}(r,\varphi)\right|^2 r\,dr\,d\varphi. \quad (14)$$

Equations (11) and (13) lead to the following prediction, which represents the main result of the analysis: Light propagating in one core of an MCF may induce nonlinear index variations in a different core, through the stimulation of guided acoustic modes. Similar to the Kerr effect, the magnitude of nonlinear index perturbations scales with the optical power of a pump wave. Unlike Kerr nonlinearity, however, the mechanism is highly frequency dependent due to phase matching considerations. Note that nonlinear opto-mechanical index changes do not require any direct coupling of optical power between the cores, and may take place even when such coupling is negligible. The magnitude of the effect is mode-dependent, through the spatial overlap integrals associated with electrostrictive stimulation and photo-elastic perturbation. Lastly, the analysis also suggests opto-mechanically induced birefringence, as $\overline{\delta\tilde{n}}^{(m,1)}(\Omega) \neq \overline{\delta\tilde{n}}^{(m,2)}(\Omega)$ even when the linear optical properties of silica are assumed to be perfectly isotropic.

The optical phase that is accumulated by the probe wave over an MCF of length $L$ would oscillate at angular frequency $\Omega$, with a magnitude that is polarization dependent:

$$\delta\tilde{\phi}_{OM}^{(m,i)}(\Omega) = k_0 \overline{\delta\tilde{n}}^{(m,i)}(\Omega) L. \quad (15)$$

Here $k_0$ is the vacuum wave-number of the probe wave, and $i = 1, 2$. The extent of such inter-core, opto-mechanical XPM is conveniently quantified in terms of an equivalent nonlinear opto-mechanical coefficient, in units of $W^{-1} \times m^{-1}$:

$$\gamma_{OM}^{(m,i)}(\Omega) = -\gamma_0^{(m,i)} \frac{1}{j + 2(\Omega - \Omega_m)/\Gamma_m}, \quad (16)$$

with maximum values on resonances:

$$\gamma_0^{(m,i)} \equiv \frac{k_0}{8n^2 c\rho_0} \frac{Q_{ES}^{(m)} Q_{PE}^{(m,i)}}{\Gamma_m \Omega_m}. \quad (17)$$

The magnitude of inter-core XPM may be expressed as:

$$\delta\tilde{\phi}_{OM}^{(m,i)}(\Omega) = \gamma_{OM}^{(m,i)}(\Omega)\tilde{P}(\Omega) L. \quad (18)$$

If the MCF is longer than the beat length of residual linear birefringence, the magnitude of phase modulation is given by the average between $\gamma_{OM}^{(m,1)}(\Omega)$ and $\gamma_{OM}^{(m,2)}(\Omega)$. The inter-core XPM in the time domain can be found by the inverse Fourier transform of $\sum_m \delta\tilde{\phi}_{OM}^{(m,i)}(\Omega)$.

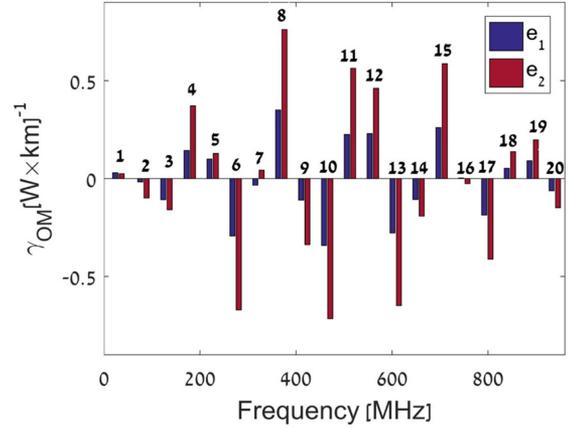

Fig. 4. Calculated nonlinear opto-mechanical coefficients between pump light at the inner core of a seven-core MCF and probe light in an outer core, as a function of the cut-off frequencies of the radial guided acoustic modes. Modal orders are noted. The cladding diameter is 125 $\mu$m, the separation between the centers of inner and outer cores is 35 $\mu$m, and the mode field diameters in all cores are 6.4 $\mu$m. Blue: probe light is polarized in the radial direction. Red: probe light is polarized in the azimuthal direction.

Depending on the acoustic mode and on the medium outside the cladding, the magnitude of the nonlinear opto-mechanical coefficient may be comparable with that of the intra-core Kerr effect, or even higher. Fig. 4 presents the calculated sets of coefficients $\gamma_0^{(m,i)}$ for a seven-core MCF of 125 $\mu$m cladding diameter that is coated with standard dual-layer acrylate coating. The linewidths $\Gamma_m$ used in the calculations were taken from previous experimental characterization [25], [30]. Pump light at the on-axis core of the MCF and a probe wave at an outer core were considered. The separation between the centers of the two cores is 35 $\mu$m, and the mode field diameters of the pump and probe fields are taken as 6.4 $\mu$m. The sign of $\gamma_0^{(m,i)}$ signifies the phase relation between the radio-frequency oscillations of the pump power and those of the local refractive index.

Inter-core XPM in the specific example is the most pronounced through mode $R_{0,8}$, with $\gamma_0^{(8,2)} = 0.76 \, \text{W}^{-1} \times \text{m}^{-1}$. By contrast, $\gamma_0^{(7,i)}$ vanish almost entirely due to poor spatial overlap in photo-elastic modulation [25]. XPM through the lowest-order modes is inefficient due to their broad transverse profiles (see also Fig. 1). For high-order modes, the efficiency is degraded by short-period spatial oscillations of the displacement profiles within the optical mode field diameter. For comparison, the coefficient of intra-core Kerr nonlinearity in the same fiber is $4 \, \text{W}^{-1} \times \text{m}^{-1}$.

Opto-mechanical inter-core XPM in a commercially available, seven-core fiber was observed in a pump-probe experiment [25]. The measurement setup was used earlier for the study of forward SBS in micro-structured fibers [19] and standard single-mode fibers [17], [30]. Full experimental detail can be found in [25]. A 30 meters-long MCF under test, coated with standard dual-layer acrylate coating, was placed within a Sagnac interferometer loop (see Fig. 5). Pump pulses of nanosecond-scale duration and Watt-level peak power were launched into the inner, on-axis core of the MCF, in one direction only. Optical bandpass filters kept the pump pulses from reaching the loop output. The pump wave stimulated the oscillations of multiple guided radial acoustic modes as discussed above.

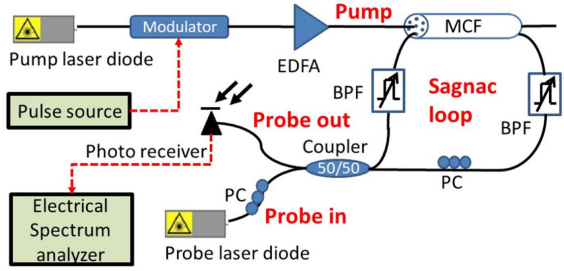

Fig. 5. Schematic illustration of the experimental setup used in the measurement of inter-core opto-mechanical cross-phase modulation in a commercially-available, seven-core fiber (MCF). PC: polarization controller; EDFA: erbium-doped fiber amplifier; BPF: optical bandpass filter.

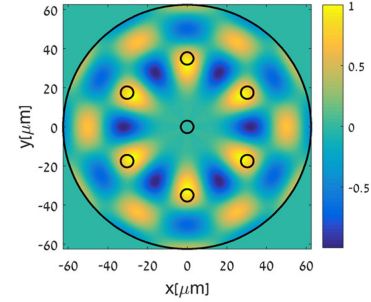

Fig. 7. Calculated normalized transverse profile of density fluctuations in the oscillations of a torsional-radial guided acoustic mode with six-fold azimuthal symmetry. The cladding diameter is 125 $\mu$m. Black circles denote the locations of the seven cores of a multi-core fiber.

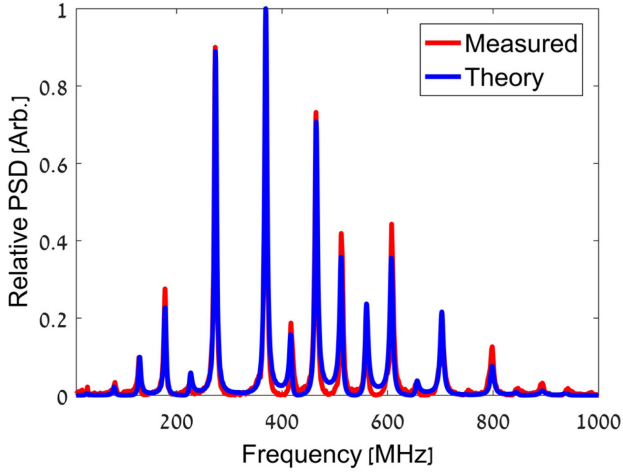

Fig. 6. Measured (red) and calculated (blue) normalized power spectral densities of opto-mechanical cross-phase modulation between a pump in the central, on-axis core of a seven-core fiber, and a probe in an outer core [25].

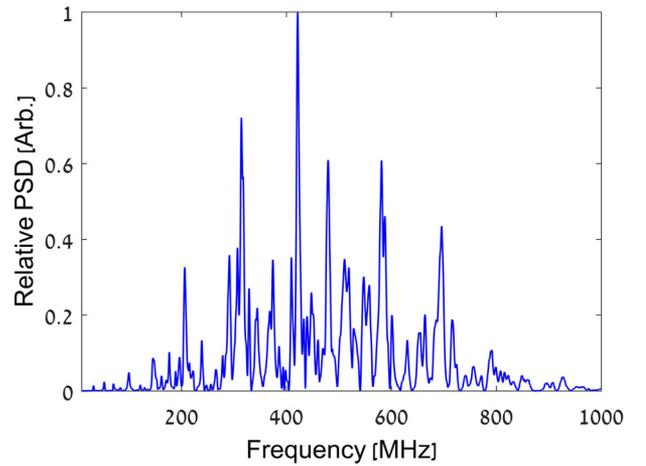

Fig. 8. Measured normalized power spectral density of opto-mechanical cross-phase modulation between pump light in the 12 o'clock off-axis outer core of a seven-core fiber, and probe light in the 8 o'clock outer core (for locations of cores see fiber cross-sections in Fig. 3 and Fig. 7).

Continuous probe light was launched into the loop in both directions: The loop passed through an outer, off-axis core of the MCF. The probe wave in the clockwise direction co-propagated with the pump pulses and acquired opto-mechanical XPM, whereas the probe in the counter-clockwise direction was largely unaffected by the acoustic waves due to lack of phase matching. The magnitude of opto-mechanical XPM was much smaller than $2\pi$ radians. The two replicas of the probe field were mixed together at the loop output. The beating converted the non-reciprocal phase modulation imposed by the acoustic waves into an intensity signal. The output probe wave was detected by a photo-receiver. For small signals, the modulation of the detector voltage was proportional to the opto-mechanical XPM of the clockwise-propagating probe wave. The voltage was monitored by a radio-frequency electrical spectrum analyzer.

Fig. 6 shows the normalized power spectral density of the measured output voltage, alongside corresponding calculations of $\sum_m |\delta\tilde{\phi}_{OM}^{(m,i)}(\Omega)|^2$ [25]. Excellent agreement is achieved between model and measurement. The results demonstrate the phase modulation of the optical probe by acoustic waves that were stimulated in a different, optically-isolated core. The XPM spectrum consists of discrete and narrow peaks, in agreement with expectations. The irregular shape of the spectrum is fully accounted for by calculations of spatial overlap. Note that the angular frequency separation between the resonances of successive modes $\Omega_{m+1} - \Omega_m$ is an order of magnitude wider than $\Gamma_{m,m+1}$. Hence XPM at a given $\Omega$ is well approximated by the contribution of one guided radial acoustic mode, at most.

If the optical pump wave is moved to an outer, off-axis core of the MCF, the radial symmetry of the electrostrictive driving force is removed. Consequently, torsional-radial guided acoustic modes of all azimuthal orders may be stimulated. Fig. 7 shows an example of the transverse profile of density fluctuations in one such mode, with six-fold azimuthal symmetry. The cut-off frequency of that mode is 223.4 MHz. The XPM spectrum between a pair of outer cores consists of contributions due to hundreds of individual modes, and becomes quasi-continuous up to several hundreds of megahertz frequencies [25]. The spectrum is qualitatively different from that of forward SBS in standard single-mode fibers. Fig. 8 shows an example of the measured spectrum of opto-mechanical XPM between two outer cores [25].

The results reported in this section show that MCFs with negligible direct coupling of optical power among the cores are not necessarily crosstalk-free. Inter-core XPM may complicate certain applications of MCFs, such as the distribution and processing of narrowband microwave signals. However, the phenomenon may also be useful. Two possible applications are presented in the next sections.

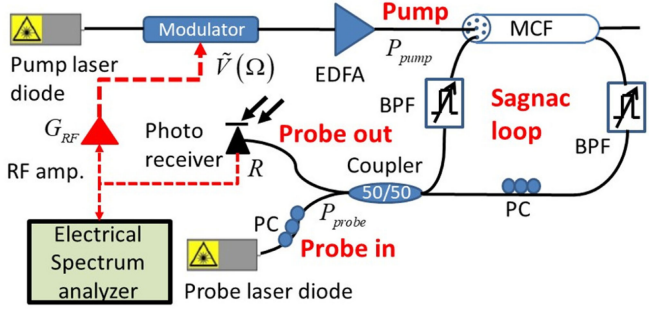

Fig. 9. Schematic illustration of an opto-electronic oscillator loop that is based on inter-core opto-mechanical cross-phase modulation in a multi-core fiber (MCF). EDFA: erbium-doped fiber amplifier; BPF: optical band-pass filter; PC: polarization controller; RF amp.: radio-frequency electrical amplifier.

## III. SINGLE-FREQUENCY OPTO-ELECTRONIC OSCILLATORS BASED ON INTER-CORE OPTO-MECHANICAL CROSS-PHASE MODULATION

Opto-electronic oscillators (OEOs) are sources of radio-frequency tones that may reach extremely low noise levels [32], [33]. The arrangement was first proposed and demonstrated by Yao and Maleki over twenty years ago [32], [33]. In the fundamental OEO configuration, a radio-frequency signal modulates continuous-wave light from an optical source, and the modulated waveform propagates over a length of fiber. The electrical signal is recovered by a photo-receiver at the output end of the fiber, and the detected voltage is amplified and fed back to drive the modulation of the input optical carrier. Given sufficient electrical gain, the hybrid opto-electronic loop may reach steady-state voltage oscillations [32], [33].

OEOs with phase noise levels as low as $-160$ dBc/Hz at 10 kHz offset have been reported in the literature [34]. However, low-noise OEOs typically involve long fiber paths, and support a large number of longitudinal modes that are closely spaced in frequency [32], [33]. The selection of a single frequency of operation requires narrowband, inline electrical band-pass filters, or other frequency discrimination mechanisms [32]–[34]. Multi-cavity OEOs over MCFs were demonstrated recently by Garcia and Gasulla [10], [35]. In 2017, we demonstrated a single-frequency OEO based on forward SBS in standard fibers [36]. Here we show that single-frequency operation may be also reached based on the frequency selectivity of inter-core XPM in MCFs [28] (see Fig. 6).

The experimental setup is illustrated in Fig. 9. It is similar to the arrangement used in the characterization of opto-mechanical inter-core XPM in Fig. 5, with one major difference: Rather than a pulse generator, the modulation of the pump wave is driven by the amplified voltage of the detected probe at the Sagnac loop output. Let us denote the average optical power of the pump wave at the MCF input as $P_{pump}$, and the amplitude of the loop voltage as $\tilde{V}(\Omega)$, with $\Omega$ the angular radio-frequency of oscillations. The pump wave passes through an electro-optic amplitude modulator that is biased at quadrature and driven by $\tilde{V}(\Omega)$. The magnitude $\tilde{P}(\Omega)$ of pump power modulation is given by $J_1[\pi \tilde{V}(\Omega)/V_\pi]P_{pump}$, where $V_\pi$ denotes the voltage difference that is necessary to switch the modulator between maximum and minimum transmission.

The modulated pump light is launched into the on-axis core of the same seven-core fiber used in the previous section, in the clockwise direction only. As discussed earlier, the pump wave induces phase modulation to a co-propagating continuous probe field in an outer core of the same fiber. We denote the optical power of the probe at the Sagnac loop input as $P_{probe}$. The magnitude of inter-core XPM is given by Eq. (18): $\delta\tilde{\theta}_{OM}(\Omega) = \sum_m \gamma_{OM}^{(m,i)}(\Omega)\tilde{P}(\Omega)L$. The state of polarization of the probe wave is aligned for maximal XPM, $i = 2$ (see Fig. 4). Note that the Kerr effect does not contribute to XPM between spatially distinct cores.

As before, the non-reciprocal XPM of the probe wave is converted to intensity modulation at the Sagnac loop output. For $\delta\tilde{\theta}_{OM}(\Omega) \ll 2\pi$, and with proper alignment of polarization, the magnitude of the probe intensity modulation is approximately $\frac{1}{2}\delta\tilde{\theta}_{OM}(\Omega)P_{probe}$. The probe wave is detected by a photo-receiver of responsivity $R$ [V $\times$ W$^{-1}$], and the recovered signal is amplified by a radio-frequency electrical amplifier of voltage gain $G_{RF}$. The electrical gain is assumed to be frequency-independent. The retrieved voltage is then fed back to drive the pump wave modulation.

Steady-state oscillations require that the sequence of electro-opto-mechanical conversions within the feedback loop should reproduce the initial modulation voltage [28]:

$$\frac{1}{2}G_{RF}RP_{probe}\sum_m \gamma_{OM}^{(m,i)}(\Omega)J_1\left[\frac{\pi\tilde{V}(\Omega)}{V_\pi}\right]LP_{pump} = \tilde{V}(\Omega). \tag{19}$$

Equation (19) may be solved to obtain the magnitude of the loop voltage as a function of pump power $P_{pump}$. Oscillations are reached at the angular frequency $\Omega$ of the longitudinal cavity mode that is closest to the opto-mechanical resonance $\Omega_m$ of the largest $|\gamma_0^{(m,i)}|$. For the MCF used in this work, that angular frequency is $\Omega_8 \approx 2\pi \times 370$ MHz. The discrete and narrow spectrum of inter-core opto-mechanical XPM therefore serves as a filter for selecting the frequency of operation. No radio-frequency electrical filters are necessary. Note that the pump and probe waves are distinct both spectrally and spatially, and the feedback loop may only close through opto-mechanical coupling.

For a small argument $\xi \ll 1$, we may approximate $J_1(\xi) \approx \frac{1}{2}\xi$, leading to the following threshold condition [28]:

$$P_{pump}P_{probe}G_{RF}R \approx \frac{4}{\pi}\frac{V_\pi}{\max\left\{\gamma_0^{(m,i)}\right\}L}. \tag{20}$$

The magnitude of the stimulated acoustic oscillations in each cross-section of the MCF is simply proportional to the loop voltage [28]:

$$\tilde{A}_r^{(m)}(\Omega) = \frac{4n}{G_{RF}RP_{probe}k_0LQ_{PE}^{(m,i)}}\tilde{V}(\Omega), \tag{21}$$

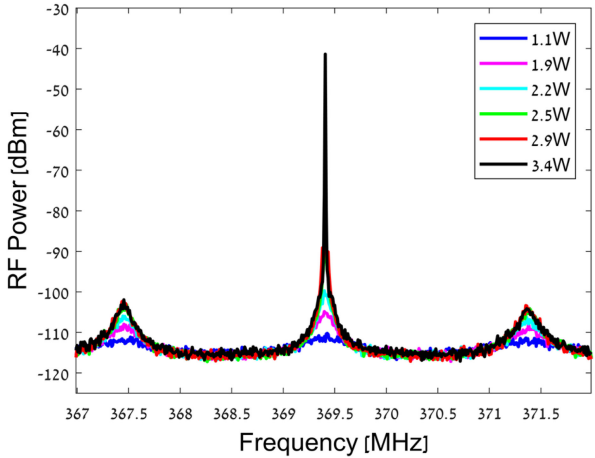

Fig. 10. Measured electrical power spectral density at the output of an opto-electronic oscillator loop that is based on opto-mechanical inter-core cross-phase modulation in a seven-core fiber [28]. Different traces correspond to different average power levels of the optical pump wave (see legend). Narrowband voltage oscillations are observed for pump power levels above a threshold of 2.5 W.

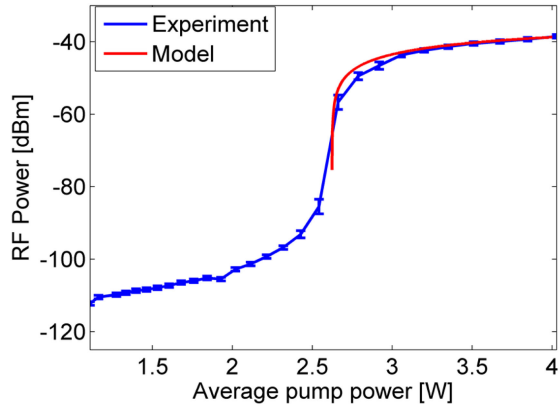

Fig. 11. Peak levels of the radio-frequency (RF) power spectral density at the output of an opto-electronic oscillator loop, as a function of the optical pump power [28]. The oscillator is based on inter-core opto-mechanical cross-phase modulation in a seven-core fiber. Blue: Experimental data. Red: Calculations of Eq. (19).

where again $m$ is the order of the radial acoustic mode for which $|\gamma_0^{(m,i)}|$ is the largest. Steady state oscillations of the loop voltage therefore signify equally narrowband, single-frequency oscillations of stimulated acoustic waves.

Fig. 10 shows measurements of the electrical power spectral density at the OEO loop output as a function of $P_{pump}$ [28]. The MCF within the Sagnac loop was 30 meters long, the probe power $P_{probe}$ was 3.7 mW, $V_\pi$ of the electro-optic modulator was 3.5 V, the detector responsivity $R$ equaled 27 V × W$^{-1}$, and the radio-frequency voltage gain $G_{RF}$ within the OEO loop was 1,200. Threshold behavior is clearly observed. For pump power well below 2.5 W, the spectra consist of multiple, comparatively broad and weak peaks that are separated by a free spectral range of about 2 MHz. The peaks represent noise that is filtered by the longitudinal modes of the OEO loop. Above threshold, narrowband oscillations appear at 369.4 MHz, the frequency of the longitudinal mode which is nearest to $\Omega_8/(2\pi)$. Fig. 11 shows the measured peak value of the output radio-frequency power spectrum as a function of pump optical power, alongside corresponding calculations of Eq. (19). General agreement is observed. The frequency of oscillations may be switched among several values by moving the pump and/or probe waves to different cores [28].

The full width at half maximum of the voltage spectrum at 3.5 W of pump power is below 100 Hz. Competing longitudinal and acoustic modes are suppressed by at least 55 dB, and harmonic distortion is below 40 dB [28]. The phase noise of the oscillator at 10 kHz offset is a modest −70 dBc/Hz, however it may be improved with a longer loop delay. The principle can be used in acoustic spectroscopy, as the frequency and magnitude of oscillations may be affected by mechanical properties of media outside the fiber (see also in the next section). Other potential applications include the generation and distribution of radio-frequency carriers, and microwave-photonic signal processing. Future work would address opto-mechanical oscillations over long MCFs with no electrical gain, and locking of multiple acoustic modes.

## IV. POINT-SENSING OF MEDIA OUTSIDE THE CLADDING OF MULTI-CORE FIBERS

Optical fibers constitute an exceptional sensing platform [37], [38]. The propagation of light in fiber may be affected by many environmental parameters such as temperature, static strain, sound and vibrations, electrical and magnetic fields, presence and concentration of chemical and biological reagents, and more [37], [38]. Optical fibers support measurements over long ranges up to hundreds of km, may be simply embedded within many structures, are comparatively immune to electromagnetic interference, and may be installed in hazardous environments where the use of electricity is prohibited [37], [38].

Alongside their many advantages, optical fiber sensors also face an inherent, fundamental challenge. Standard optical sensing protocols, such as the measurements of absorption, refraction, scattering or fluorescence, mandate spatial overlap between light and the substance under test. Standard optical fibers, on the other hand, are designed to do the opposite: guide light at inner cores, and reduce any leakage of light to the outside to the absolute minimum. Consequently, many optical fiber sensors monitor parameters that prevail within the cores, such as temperature or axial strain. Those sensors that monitor chemical species often rely on non-standard micro-structured geometries, specialty coatings, non-standard reactive materials, and/or considerable structural modifications of standard fibers [39]–[44].

Forward SBS processes provide a way around this difficulty. As discussed in Section II, the transverse profiles of guided acoustic modes span the entire cladding cross-section, and reach the outer cladding boundary (see also Fig. 1). The acoustic oscillations are therefore affected by the properties of materials outside the cladding. In particular, the modal linewidths $\Gamma_m$ are broadened by the dissipation of acoustic energy to the surrounding medium. The linewidths are approximately given by [30]:

$$\Gamma_m \approx \Gamma_m^{(\text{int})} + \frac{v_d}{a} \ln\left(\left|\frac{Z_{SiO2} + Z_{out}}{Z_{SiO2} - Z_{out}}\right|\right). \quad (22)$$

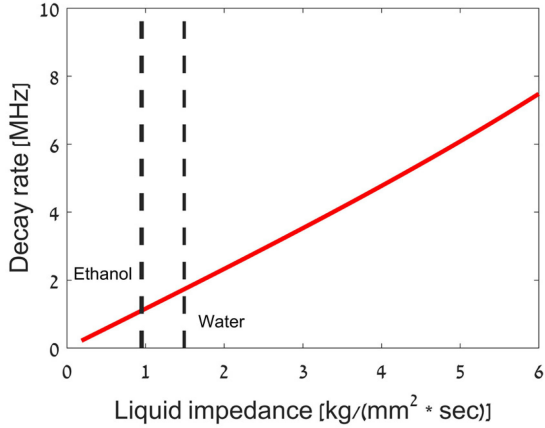

Fig. 12. Decay rates (half the linewidths) of guided acoustic modes in a silica optical fiber as a function of the mechanical impedance of the substance outside the cladding. The mechanical impedances of water and ethanol are noted in dashed black lines.

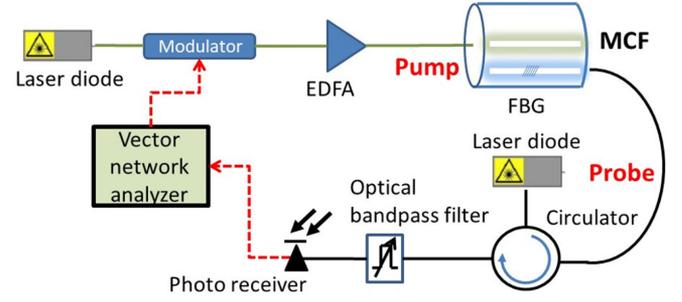

Fig. 13. Schematic illustration of the experimental setup used in opto-mechanical point-sensing of media outside the cladding of a seven-core fiber (MCF) [29]. Pump light is amplitude-modulated by the output voltage of a radio-frequency vector network analyzer and launched into the central, on-axis core of the fiber. A continuous probe wave is propagated in an outer core of the fiber in the opposite direction. The probe wave is reflected from a fiber Bragg grating (FBG) that is inscribed in the same outer core. The reflected probe wave is detected by a photo-receiver, and amplitude modulation of the obtained signal is monitored by the input port of the vector network analyzer. EDFA: erbium-doped fiber amplifier.

Here $\Gamma_m^{(\text{int})}$ represents the linewidth due to internal acoustic dissipation in silica. This component is mode-specific and does not vary with changes in surrounding media. For $R_{0,8}$, $\Gamma_8^{(\text{int})} \approx 2\pi \times 100$ kHz [30]. Also in Eq. (22), $Z_{SiO2}$ is the mechanical impedance of silica, and $Z_{out}$ is the unknown impedance of the medium outside the cladding. Measurements of opto-mechanical cross-phase modulation spectra may therefore retrieve $Z_{out}$ (see Fig. 12). The internal linewidths components $\Gamma_m^{(\text{int})}$ can be pre-calibrated through measurements in which a bare fiber is kept in air ($Z_{out} \ll Z_{SiO2}$).

The prospect of opto-mechanical sensing outside the cladding has led to renewed interest in forward SBS processes in standard fibers [30], [45]–[51]. Previous works reported position-integrated measurements of liquids that were accumulated over the entire lengths of fibers under test [30], [45]–[47], [50], and distributed sensing with spatial resolution of 15-50 meters [48], [49], [51]. However, the most basic and widely employed category of optical fiber sensors is point-measurement of a parameter of interest. Point sensing based on the fiber opto-mechanics remained missing. Here we show that inter-core opto-mechanical XPM in MCFs can address this challenge as well [29].

Similarly to previous arrangements, the sensor is based on modulated pump light that is propagating in the central, on-axis core of a seven-core fiber. The pump wave induces index modulation in the outer cores of the MCF. The magnitude of index perturbation at position $z$ along the fiber, $\overline{\delta\tilde{n}}^{(m,i)}(\Omega, z)$, is given by Eq. (11) through Eq. (14). The photo-elastic perturbations may vary with position due to changes in the local surrounding media. A fiber Bragg grating (FBG) with maximum power reflectivity $R_{\max}$ and reflectivity spectrum of full width at half maximum $\Delta\omega_{FBG}$ is inscribed in an outer core at position $z_{FBG}$. A continuous optical probe wave of optical power $P_{probe}$ is propagated in the same core. The optical angular frequency $\omega_0$ of the probe wave is tuned to a spectral slope of the FBG reflectivity spectrum.

Fiber opto-mechanics affects the reflectivity of the probe wave through two mechanisms. First, the FBG transfer function converts any phase modulation of the probe to intensity variations. Second, photo-elastic perturbations to the FBG itself modify its transfer function, and give rise to additional intensity modulation of the reflected probe. The modulation depth of the reflected probe wave is given by:

$$\beta(\Omega) \approx \frac{\delta\tilde{P}_{probe}(\Omega)}{\frac{1}{2}R_{\max}P_{probe}} \approx 2Q_{FBG}$$
$$\times \left[ j\frac{\Omega}{c} \int_{z_S}^{z_F} \overline{\delta\tilde{n}}^{(m,i)}(\Omega, z)\,dz + \frac{\overline{\delta\tilde{n}}^{(m,i)}(\Omega, z_{FBG})}{n} \right]. \quad (23)$$

Here $\delta\tilde{P}_{probe}(\Omega)$ denotes the modulation magnitude of the reflected probe power, $Q_{FBG} \equiv \omega_0/\Delta\omega_{FBG}$ is the quality factor of the FBG transfer function, and $z_{S,F}$ are the start and end points of the fiber section over which opto-mechanical XPM is accumulated prior to reflection. The first term in the intensity modulation of the reflected probe wave represents the position-integrated effect of inter-core XPM, whereas the second term is strictly local. Unlike many applications of FBGs in fiber sensing, the axial period of the grating is unaffected.

The relative magnitude of the two terms in Eq. (23) may be evaluated for two cases. When the probe wave is co-propagating with the pump, opto-mechanical inter-core XPM accumulates over the entire physical length of the MCF, from the point of entry to $z_{FBG}$. If that length exceeds tens of centimeters, the position-integrated term in Eq. (23) dominates over the local contribution to the modulation of the reflected probe power. Localized opto-mechanical sensing of substances outside the cladding would be difficult to achieve in this case.

On the other hand, if the probe wave is counter-propagating with respect to the pump, the lack of phase matching would lead to complete cancellation of opto-mechanical inter-core XPM following a propagation distance $\Lambda_z = \pi c/(n\Omega)$. With proper choice of $z_{FBG}$, the effective distance of XPM accumulation $|z_F - z_S|$ can be very short. In this case, power modulation of the reflected probe wave would be dominated by the local magnitude of photo-elastic index changes at $z_{FBG}$. Note, however, that the residual position-integrated term of Eq. (23) for a mismatched, counter-propagating probe wave may still be nonzero, and could overshadow the local contribution, depending on $z_{FBG}$.

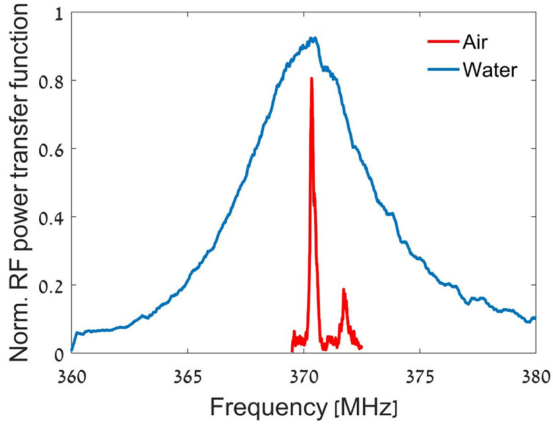

Fig. 14. Measured normalized radio-frequency (RF) power transfer functions between the input modulation of a pump wave at the inner core, and the modulation of a probe wave that was reflected from a fiber Bragg grating in an outer core [29]. The pump and probe waves were launched from opposite ends of the fiber under test. Red– The uncoated fiber section containing the short grating was kept in air. Blue–The same fiber section was immersed in water.

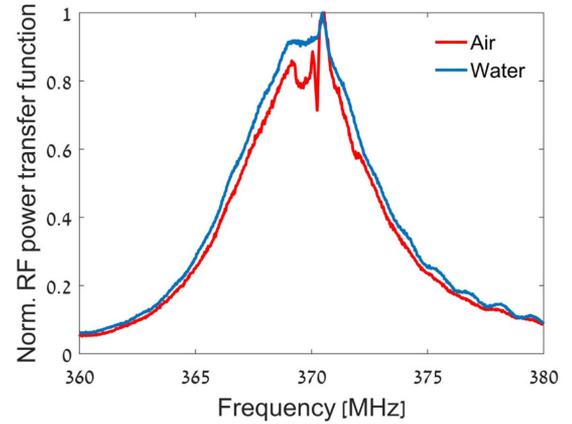

Fig. 15. Measured normalized radio-frequency (RF) power transfer functions between the input modulation of a pump wave at the inner core, and the modulation of a probe wave that was reflected from a fiber Bragg grating in an outer core [29]. The pump and probe waves were launched from the same end of the fiber under test. Red–The uncoated fiber section containing the short grating was kept in air. Blue–the same fiber section was immersed in water.

With scanning of the angular frequency $\Omega$, the local spectrum of forward SBS may be evaluated, leading to point-sensing of the mechanical impedance of the local medium outside the cladding at $z_{FBG}$. For the MCF and FBGs used in this work, air outside the cladding and pump power of 1 W, the photo-elastic perturbations are rather weak: $\overline{\overline{\delta n}}^{(m,i)}(\Omega_8, z_{FBG})$ is on the order of 1e-9 refractive index units, and the probe modulation depth $\beta(\Omega_8)$ is on ten ppm scale (see Eq. (23)). The effect is weaker still when the MCF is coated or immersed in liquid. However, since the probe wave is separated from the pump both spatially and spectrally, and the modulation bandwidth in each measurement is very narrow, background noise is very low and weak signals may still be detected.

The setup is illustrated in Fig. 13 [29]. Measurements were carried out over a 30 meters-long segment of the same MCF used in previous sections. The entire length of the fiber was coated with standard dual-layer acrylate coating. An FBG was inscribed in one of the outer cores, at a distance $z_{FBG}$ of 20 meters from the input end of the pump wave. The peak reflectivity wavelength of the FBG was 1572 nm, and the full width at half maximum of its reflectivity spectrum was 0.15 nm (for the inscription of FBGs in MCFs, see [52]). The coating was removed from a 4 cm-long section of the fiber, containing the FBG. The length of the grating itself was 0.8 cm.

Pump light at 1542 nm wavelength was amplitude-modulated by a sine wave of variable angular frequency $\Omega$, from the output port of a radio-frequency vector network analyzer. The pump wave was amplified to an average power of 32 dBm by an erbium-doped fiber amplifier, and was launched into the inner core of the MCF. A continuous probe wave at 1571.9 nm wavelength was propagated in the outer core of the MCF that contained the FBG. The probe wavelength was within the slope of the reflectivity spectrum of the FBG. The input power of the probe wave was 0 dBm. Reflections of the probe wave were routed by a fiber-optic circulator and detected by a photo-receiver with a rise time of 15 ps. An optical bandpass filter was used to keep residual leakage of the pump wave from reaching the detector. Intensity modulation of the reflected probe wave was measured using the input port of the vector network analyzer. The value of $\Omega$ was scanned over a range of $2\pi \times 20$ MHz around $\Omega_8$.

Fig. 14 shows the measured normalzized radio-frequency power transfer functions between the input modulation of the pump wave and the output modulation of a counter-propagating probe wave, following reflection from the FBG [29]. The exposed fiber section containing the FBG was kept in air (red), or immersed in water (blue). Resonant modulation at 370.3 MHz frequency was observed for the FBG in air, in agreement with the known value of $\Omega_8/(2\pi)$. The modulation spectrum was very narrow, with a full width at half maximum of only 220 kHz. The spectrum was broadened when the exposed fiber section was immersed in water instead. The measurements clearly distinguish between the two media outside the fiber cladding. While the resonance frequency could still be identified with the FBG in water, the peak was much weaker and the spectral width was broadened to 6 MHz. The observed bandwidth is broader than $\Gamma_8/(2\pi)$ in a bare fiber in water (see also Fig. 12). The observed forward SBS spectrum was likely dominated by residual, position-integrated phase modulation background of the mismatched, counter-propagating probe wave, over the fiber length leading to the FBG. The local forward SBS spectrum of the exposed fiber section in water could not be resolved below this background.

Fig. 15 shows corresponding traces of normalized radio-frequency power transfer functions, obtained for a probe wave that was co-propagating with the pump [29]. Results for the FBG in air and in water are shown in red and blue traces, respectively. A resonant frequency of 369.8 MHz is observed in both traces.

The full width at half maximum of both transfer functions was 6.3 MHz. The width matches that of F-SBS through mode $R_{0,8}$ in a coated fiber [30], and it is unaffected by the local medium outside the cladding at $z_{FBG}$. A narrow spectral dip near the resonance frequency could be observed when the FBG was kept in air. This feature corresponds to the local photo-elastic perturbation of the grating itself, which was also seen in Fig. 14. This local perturbation is overlaid on top of the position-integrated F-SBS spectrum of the coated fiber. The contribution of the local F-SBS spectrum was weaker with the FBG in water. The transfer function of a reflected, co-propagating probe provides a position-integrated measurement of forward SBS that is free of interferometric arrangements. The use of FBGs presents a simpler alternative to the Sagnac loop in position-integrated measurements of fiber opto-mechanics.

The results constitute the first forward SBS sensing experiment involving MCFs, and illustrate the value of such fibers towards sensing applications. The proposed concept represents a significant step towards the establishment of forward SBS as a complete sensor platform, supporting position-integrated, spatially-distributed as well point measurements. The measurement sensitivity was restricted by accumulation of residual position-integrated cross-phase modulation of the counter-propagating probe wave, as noted above. This background may be removed if the position of the readout fiber Bragg grating is chosen judiciously. However, such studies are outside the scope of the present work. The technique is also applicable to the monitoring of forward SBS with FBGs that are re-coated with a proper material, such as a thin layer of polyimide [50], [51].

## V. CONCLUSION

MCFs provide a rich playground for the study of opto-mechanics, due to the plurality of guided optical and acoustic modes. In this work, we have summarized our studies of opto-mechanical coupling in MCFs of the last three years. Analysis provides the following primary prediction: the stimulation of guided acoustic waves may give rise to inter-core opto-mechanical XPM, even when the direct coupling of optical power among the cores is negligible. This result has been demonstrated experimentally, in good quantitative agreement with model predictions. The XPM spectrum due to radial guided acoustic modes consists of discrete and narrow peaks. The propagation of light in outer, off-axis cores of an MCF may give rise to quasi-continuous inter-core XPM spectra, through the stimulation of more general, torsional-radial guided acoustic modes.

The magnitude of the effect is quantified in terms of an equivalent nonlinear opto-mechanical coefficient, which varies strongly with the acoustic frequency and depends on the choices of cores and states of polarization. The strength of the effect on resonance may be comparable with that of intra-core XPM through the Kerr effect: the equivalent nonlinear coefficient in coated fibers is on the order of $1~W^{-1} \times km^{-1}$. The pump power levels that are necessary for opto-mechanical inter-core XPM applications are on the order of hundreds of mW to few Watts, depending on length.

Inter-core XPM is expected to have little effect on space-division multiplexing optical communication networks. Communication data spread over tens of GHz, and only a small fraction of their optical power modulation spectra falls within the narrow bandwidths of acoustic waves stimulation. The effect can become more significant in the distribution of narrowband signals, such as in sub-carrier multiplexed networks or radio-over-fiber links [53]. Radio frequency modulation at specific frequencies may be coupled among cores.

We have proposed and demonstrated two potential applications of MCF opto-mechanics. In a first example, inter-core XPM was used to close an electro-opto-mechanical feedback loop which passed through the MCF [28]. The discrete and narrow XPM spectrum provided the frequency discrimination that is necessary for single-frequency OEO operation. Inline electrical filters were not necessary. The loop voltage oscillations were accompanied by the stimulation of highly coherent acoustic waves in every MCF cross-section. The concept is applicable to integrated-photonic OEOs with multiple, isolated optical waveguides that are coupled mechanically. Point-sensing of media outside the MCF cladding was demonstrated as well [29]. The results complement previous reports of position-integrated and spatially-distributed measurements [30], [45]–[51], and provide a significant piece that was missing in the new toolset of opto-mechanical fiber sensors.

Inter-core opto-mechanical crosstalk plays a role in the collective behavior of multiple optical modes in a fiber, with or without direct optical coupling. Studies of the effect have just begun. Future works would address further implications of the opto-mechanical interactions in MCFs on communication, microwave photonics, sensing and fiber laser systems.

**Hilel Hagai Diamandi** received the B.Sc. and M.Sc. degrees in electrical engineering from Bar-Ilan University, Ramat-Gan, Israel, in 2015 and 2017, respectively. His M.Sc. research was dedicated to opto-mechanical effects in multicore fibers. He is currently working toward the Ph.D. degree in electrical engineering with Bar-Ilan University.

His research interests include optical fiber sensors, and opto-mechanical interactions in optical fibers. He received the Bar-Ilan University Rector Award for excellence in undergraduate studies twice: in 2013 and 2015, and the same Award for excellence in graduate studies in 2017. In 2018, he was awarded the Azrieli Fellowship for doctoral studies by the Azrieli Foundation.

**Yosef London** received the B.Sc. and M.Sc. degrees in physics and electrical engineering from Bar-Ilan University, Ramat-Gan, Israel, in 2013 and 2015, respectively. He is currently working toward the Ph.D. degree in electrical engineering with Bar-Ilan University.

His research interests include optical fiber sensors, nonlinear optics, and opto-mechanical interactions in fibers and photonic integrated circuits. He received the Bar-Ilan University Rector Award for excellence in graduate studies in 2015.

**Arik Bergman** (M'17) received the B.Sc., M.Sc., and Ph.D. degrees in electrical engineering from Tel-Aviv University, Tel Aviv-Yafo, Israel, in 2006, 2011, and 2017, respectively.

Since 2016, he has been a Postdoctoral Research Fellow with the Faculty of Engineering, Bar Ilan University, Ramat-Gan, Israel. Since 2018, he has also been affiliated with the City University of New York, New York, NY, USA. His research interests include opto-mechanical interactions in integrated photonic devices and fibers, Brillouin lasers, phononic oscillators, and fiber imaging and sensor systems.

**Gil Bashan** received the B.Sc. and M.Sc. degrees in physics and electrical engineering from Bar-Ilan University, Ramat-Gan, Israel, in 2016 and 2018, respectively. He is currently working toward the Ph.D. degree in electrical engineering with Bar-Ilan University.

His research interests include optical fiber propagation effects, opto-mechanics, and Brillouin scattering. He received the Bar-Ilan University Rector Award for excellence in graduate studies in 2018.

**Javier Madrigal** received the M.Sc. degree in telecommunications engineering from Universitat Politécnica de Valéncia, Valencia, Spain, in 2015. He is currently working toward the Ph.D. degree in communications engineering at the same university.

Since 2015, he has been working as a Researcher with the Photonics Research Labs, iTEAM Research Institute, Valéncia, Spain, where he has been involved in the design and fabrication of fiber Bragg gratings. His research interests include multicore fibers, optical fiber sensors, long period gratings, and tilted fiber Bragg gratings.

**David Barrera** received the M.Sc. and Ph.D. degrees from the Universitat Politécnica de Valéncia, Valencia, Spain, in 2008 and 2013, respectively.

He is currently a Postdoctoral Researcher with the Universidad de Alcalá, Alcalá de Henares, Spain. His current research interests include optical fiber sensors, fiber Bragg gratings, and multicore optical fibers.

**Salvador Sales** (S'93–M'98–SM'04) received the M.Sc. and Ph.D. degrees from the Universitat Politécnica de Valéncia, Valencia, Spain. Since 2007, he has been a Professor with the iTEAM Research Institute, Universitat Politécnica de Valéncia.

He is a coauthor of more than 125 journal papers and 300 international conference papers. He has been collaborating and leading several national and European research projects since 1997. His main research interests include optoelectronic signal processing for optronic and microwave systems, optical delay lines, fiber Bragg gratings, WDM and SCM lightwave systems, and semiconductor optical amplifiers.

Dr. Sales was the recipient of the Annual Award of the Spanish Telecommunication Engineering Association for the best Ph.D. on optical communication.

**Avi Zadok** received the B.Sc. degree in physics and mathematics from the Hebrew University of Jerusalem, Jerusalem, Israel, in 1994, the M.Sc. degree in physical electronics from Tel-Aviv University, Tel Aviv-Yafo, Israel, in 1999, and the Ph.D. degree in electrical engineering from the same University in 2007.

Between 2007 and 2009, he was a Postdoctoral Research Fellow with the Department of Applied Physics, California Institute of Technology. In 2009, he joined the Faculty of Engineering, Bar-Ilan University, Ramat-Gan, Israel, where he has been a Full Professor since 2017. He is the Chairman of the Israel Young Academy in 2019–2020. He is the coauthor of 140 papers in scientific journals and proceedings of international conferences. His research interests include fiber-optics, nonlinear optics, integrated photonic devices, and opto-mechanics.

Dr. Zadok received the Krill Award of the Wolf Foundation in 2013. He received a Starter Grant from the European Research Council (ERC) in 2015.